\documentclass[prc,twocolumn,showpacs,preprintnumbers,amsmath,amssymb]{revtex4}
\usepackage{graphicx}% Include figure files
\usepackage{dcolumn}% Align table columns on decimal point
\usepackage{bm}% bold math

\begin{document}

%\preprint{}

\title{Charge-dependent calculations of single-particle energies in nuclei
around $^{16}$O with modern nucleon-nucleon interactions}

% Force line breaks with \\

\author{Shinichiro Fujii}
 \email{sfujii@nt.phys.s.u-tokyo.ac.jp}
\affiliation{%
Department of Physics, University of Tokyo, Tokyo 113-0033, Japan
%RI Beam Factory Project Office,
%RIKEN (The Institute of Physical and Chemical Research),
%Wako 351-0198, Japan
%This line break forced with \textbackslash\textbackslash
}%
\author{Ryoji Okamoto}
% \email{okamoto@mns.kyutech.ac.jp}
\author{Kenji Suzuki}
% \email{suzuki@mns.kyutech.ac.jp}
\affiliation{
Department of Physics, Kyushu Institute of Technology,
Kitakyushu 804-8550, Japan
}

\date{\today}% It is always \today, today,
             %  but any date may be explicitly specified

\begin{abstract}
The binding energies of the ground states and several excited states related to
single-particle and -hole states in nuclei around $^{16}$O are
calculated taking charge dependence into account.
Effective interactions on the particle basis
are constructed from modern charge-dependent nucleon-nucleon interactions
and the Coulomb force within the framework of the unitary-model-operator
approach.
Single-particle (-hole) energies are obtained from the energy differences of
the binding energies between a particle (hole) state in $^{17}$O or $^{17}$F
($^{15}$N or $^{15}$O) and the ground state of $^{16}$O.
The resultant spin-orbit splittings are small for the hole state and large
for the particle state in comparison with the experimental values
though the differences between the experimental and calculated values are
not very large.
The charge dependence of the calculated single-particle energies
for the ground states are in good agreement with the experimental values.
Furthermore, the Thomas-Ehrman shift due to the Coulomb force
for the 1$s_{1/2}$ states in $^{17}$O and $^{17}$F can be observed.

\end{abstract}

% PACS, the Physics and Astronomy
                             % Classification Scheme.
%\keywords{Suggested keywords}%Use showkeys class option if keyword
                              %display desired
\maketitle

\section{\label{sec:introduction}Introduction}

The single-particle level is one of the fundamental structures in nuclei.
Important physical quantities such as the spin-orbit splittings
and the magic numbers are characterized by the single-particle level.
Recently, it has been argued that some magic numbers disappear
and new magic numbers arise in nuclei
near the drip lines~\cite{Ozawa00,Otsuka01}.
When we calculate the energies of single-particle levels in neutron- or
proton-rich nuclei, it would be desirable that the calculation formalism
is based on the particle basis.
Advantages of the particle-basis formalism are that the
Coulomb force can be treated accurately for the proton-proton channel and
effects of charge dependence in realistic nuclear forces
are taken into account in structure calculations.
In the particle-basis formalism, one can obtain the energy
differences between proton and neutron levels for not only $N=Z$
nuclei but also neutron- or proton-rich nuclei in the same manner.

The calculation of single-particle energies starting with
a nucleon-nucleon force in free space is a fundamental problem in theoretical
nuclear physics.
There have been many attempts to understand the structure of
single-particle levels as well as other ground-state properties
in nuclei~\cite{Kuemmel78,Ando81,Zamick92,Pieper93,Muether00}.
In such calculations,
we need a many-body theory that leads to an effective interaction
in a restricted model space for a nucleus in many cases.
For this purpose, the $G$ matrix has been widely used as a basic ingredient
in performing structure calculations~\cite{Kuo90,Hjorth-Jensen95}.
A recent study of the doubly closed-shell nucleus $^{16}$O
using the $G$ matrices
constructed from modern nucleon-nucleon interactions can be seen
in Ref.~\cite{Gad02}.
Lately, as an alternative of the $G$ matrix, a low-momentum potential
$V_{{\rm low}-k}$ has been constructed from a realistic nucleon-nucleon
interaction using a renormalization group technique or conventional
effective interaction theory by Bogner {\it et al}~\cite{Bogner03}.
The application of $V_{{\rm low}-k}$ to the calculation of ground-state
properties of the closed-shell nuclei $^{16}$O and $^{40}$Ca has been done
in Ref.~\cite{Coraggio03}.

As one of the methods for solving nuclear many-body problems,
we have developed the unitary-model-operator approach (UMOA)~\cite{Suzuki94}.
An energy-independent and Hermitian effective interaction is derived
through a unitary transformation of an original Hamiltonian.
Nuclear ground-state properties, such as the ground-state energy,
charge radius, and single-particle energy have been calculated for
$^{16}$O~\cite{Suzuki87} and $^{40}$Ca~\cite{Kumagai97}.
We have learned that spin-orbit splittings for hole states
are enlarged by taking second-order diagrams into account.
Furthermore, the UMOA has been developed for the structure calculation
of $\Lambda$ hypernuclei
and applied to $^{16}_{\Lambda}$O, $^{17}_{\Lambda}$O,
and $^{41}_{\Lambda}$Ca using hyperon-nucleon interactions
in free space~\cite{Fujii00,Fujii02}.
Differences in the properties of modern hyperon-nucleon interactions
have been disclosed in the structure calculation.

Recently, we have extended the formulation of the UMOA from the isospin basis
to the particle one for the purpose of the charge-dependent calculation.
To confirm the validity of the calculation method based on the particle basis,
in this paper, we apply this method to
$^{16}$O and its neighboring nuclei $^{15}$N, $^{15}$O, $^{17}$O, and $^{17}$F.
Binding energies of these nuclei are calculated
for the ground states, and excited states which
have the single-particle or single-hole structure as the main component.
The single-particle energy in the neighboring nuclei
is given as the relative energy between a single-particle (-hole) state
in the neighboring nuclei and the ground state of $^{16}$O.
As for the single-particle (-hole) state,
the excitation up to two-particle one-hole
(one-particle two-hole) from the unperturbed ground state of $^{16}$O
are taken into account.

Four high-precision nucleon-nucleon interactions represented
in momentum space are employed,
namely, the Nijmegen~93 (Nijm~93)~\cite{Stoks94}, Nijm~I~\cite{Nijmegen},
the charge-dependent Bonn (CD~Bonn)~\cite{Machleidt96},
and the next-to-next-to-next-to-leading order (N$^{3}$LO)
potential~\cite{Entem03} based on
chiral perturbation theory~\cite{Weinberg90,Ordonez94} which has recently
been constructed by Entem and Machleidt.
In these potentials, effects of charge dependence are taken into account.
The first three potentials are based on
meson-exchange models in which several kinds of meson are incorporated.
On the other hand, the essential degrees of freedom of the mesons
in the N$^{3}$LO potential are only for the pions.
Therefore, the N$^{3}$LO potential is constructed in
a low-momentum region compared to
the meson-exchange potentials which have heavier mesons.
However, the N$^{3}$LO potential has the high accuracy to reproduce
the nucleon-nucleon data below $E_{\rm lab}=290$ MeV, and thus
the N$^{3}$LO potential as well as other high-precision nucleon-nucleon
interactions can be used in nuclear structure calculations.

This paper is organized as follows.
In Sec.~\ref{sec:method}, the methods for deriving effective interactions
and performing structure calculations are given.
In Sec.~\ref{sec:results}, calculated results for $^{16}$O and its neighboring
nuclei using the four realistic interactions are presented.
Finally, we summarize the present work in Sec.~\ref{sec:summary}.

\section{\label{sec:method}Method of calculation}

In the UMOA, the Hamiltonian to be considered is given by a cluster expansion
of a unitarily transformed Hamiltonian.
In the previous works~\cite{Suzuki94,Suzuki87}, many-body correlations up to
three-body cluster terms have been evaluated.
It has been confirmed that the cluster expansion in the numerical calculation
for $^{16}$O shows the good convergence at the three-body cluster level.
We may say that since we consider the $N\simeq Z$ nuclei around $^{16}$O
in the present study, the three-body cluster terms do not have a significant
contribution to the energy difference between the single-particle (-hole)
levels of the proton and neutron.
Therefore, in the present calculation,
we neglect the three-body cluster terms for simplicity.
The evaluation of the three-body cluster terms based on the particle basis
is a further challenge which should be accomplished
for a deeper understanding of nuclei.

In the following subsections, we present a general framework
for deriving an effective interaction and a practical method for
the structure calculation in the present study.

\subsection{Derivation of effective interaction in the $P$ and $Q$ spaces}

In the usual sense of effective interaction theory, an effective interaction
is defined in a low-momentum model space ($P$ space).
However, in general, one can also derive an effective interaction
in the complement ($Q$ space) of the $P$ space by making the decoupling
condition for the effective interaction $\tilde{v}$ as $Q\tilde{v}P=0$.
Note that the projection operators $P$ and $Q$ satisfy the usual relations as
$P+Q=1$, $P^{2}=P$, $Q^{2}=Q$, and $PQ=QP=0$.
We here present a general framework for deriving an two-body effective
interaction of Hermitian type for a two-body system.

The two-body effective interaction $\tilde{v}_{12}$ of Hermitian type
is written as
\begin{equation}
\label{eq:eff_int}
\tilde{v}_{12}=U^{-1}(h_{0}+v_{12})U-h_{0},
\end{equation}
where $v_{12}$ is the bare two-body interaction and
$h_{0}$ is the one-body part of the two-body system which consists of
the kinetic energy $t_{1}$ ($t_{2}$) and, if necessary,
the single-particle potential $u_{1}$
($u_{2}$) as $h_{0}=t_{1}+u_{1}+t_{2}+u_{2}$.
The operator $U$ for the unitary transformation of $h_{0}+v_{12}$
can be written as~\cite{Suzuki82}
\begin{equation}
\label{eq:U}
U=(1+\omega-\omega ^{\dagger})
(1+\omega \omega ^{\dagger} +\omega ^{\dagger}\omega )^{-1/2}
\end{equation}
by introducing the operator $\omega$ satisfying $\omega =Q\omega P$ and thus
$\omega ^{2}={\omega ^{\dagger}}^{2}=0$.
The above expression of $U$ agrees with the block form using the projection
operators $P$ and $Q$ of \=Okubo ~\cite{Okubo54} given by
\begin{equation}
\label{eq:U_block}
U=\left(
  \begin{array}{cc}
       P(1+\omega ^{\dagger}\omega)^{-1/2}P
    & -P\omega ^{\dagger}(1+\omega \omega ^{\dagger})^{-1/2}Q \\
       Q\omega (1+\omega ^{\dagger}\omega )^{-1/2}P
    &  Q(1+\omega \omega ^{\dagger})^{-1/2}Q
  \end{array}
\right).
\end{equation}
We should note here that the operator $U$ is also expressed as
\begin{equation}
\label{eq:EtoS}
U=e^{S},
\end{equation}
where $S$ is anti-Hermitian and given under the restrictive conditions
$PSP=QSQ=0$ by
\begin{equation}
\label{S}
S={\rm arctanh}(\omega -\omega^{\dagger}).
\end{equation}

In order to obtain the matrix elements of $\omega$, we first solve exactly the
two-body eigenvalue equation as
\begin{equation}
\label{eq:eigenvalue_eq}
(h_{0}+v_{12})|\Phi _{k}\rangle =E_{k}|\Phi _{k}\rangle .
\end{equation}
With the eigenvector $|\Phi _{k}\rangle$, the matrix elements of $\omega$
on the basis states $|p\rangle$ in the $P$ space and $|q\rangle$
in the $Q$ space can be determined as
\begin{equation}
\label{eq:omega_qp}
\langle q|\omega|p\rangle =\sum_{k=1}^{d}\langle q|Q|\Phi _{k}\rangle
\langle \tilde{\phi}_{k}|p\rangle,
\end{equation}
where $d$ is the dimension of the $P$ space,
and $\langle \tilde{\phi}_{k}|$ is the biorthogonal state of
$|\phi _{k}\rangle =P|\Phi _{k}\rangle$, which means the matrix inversion
$[\langle \tilde{\phi}_{k}|p\rangle ] =[ \langle p'|\phi _{k}\rangle ]^{-1}$
and satisfies
$\sum_{p}\langle \tilde{\phi} _{k}|p\rangle \langle p|\phi _{k'}\rangle
=\delta _{k,k'}$
and
$\sum_{k}\langle p'|\tilde{\phi} _{k}\rangle \langle \phi _{k}|p\rangle
=\delta _{p,p'}$.
It should be noted that the set of eigenstates
$\{ |\Phi _{k}\rangle , k=1, 2,\cdot \cdot \cdot, d\}$ is selected
so that they have the largest $P$-space overlaps among all the eigenstates
in Eq.~(\ref{eq:eigenvalue_eq}).

Then, in order to obtain the matrix elements of $U$,
we introduce the eigenvalue
equation for $\omega ^{\dagger}\omega$ in the $P$ space as
\begin{equation}
\label{eq:omega2}
\omega ^{\dagger}\omega|\alpha _{k}\rangle
=\mu _{k}^{2}|\alpha _{k} \rangle .
\end{equation}
Using the solutions to the above equation, we define
the ket vector $|\nu _{k}\rangle$ as
\begin{equation}
\label{nu_k}
|\nu _{k}\rangle =\frac{1}{\mu _{k}}\omega |\alpha _{k}\rangle ,
\end{equation}
which is also written as
\begin{equation}
\label{eq:nu_k_mat}
\langle q|\nu _{k}\rangle =\frac{1}{\mu _{k}}
\sum_{p}\langle q|\omega |p\rangle \langle p|\alpha _{k}\rangle .
\end{equation}
Using Eqs.~(\ref{eq:omega2})-(\ref{eq:nu_k_mat}),
we obtain the matrix elements of the unitary-transformation
operator $U$ in Eq.~(\ref{eq:U}) as
\begin{eqnarray}
\label{eq:U_p'p}
\langle p'|U|p\rangle
&=&\langle p'|(1+\omega^{\dagger}\omega )^{-1/2}|p\rangle \nonumber \\
&=&\sum_{k=1}^{d}(1+\mu_{k}^{2})^{-1/2}
\langle p'|\alpha _{k}\rangle \langle \alpha _{k}|p\rangle ,
\end{eqnarray}
\begin{eqnarray}
\label{eq:U_qp}
\langle q|U|p\rangle
&=&\langle q|\omega (1+\omega^{\dagger}\omega )^{-1/2}|p\rangle \nonumber \\
&=&\sum_{k=1}^{d}(1+\mu_{k}^{2})^{-1/2}\mu _{k}
\langle q|\nu _{k}\rangle \langle \alpha _{k}|p\rangle ,
\end{eqnarray}
\begin{eqnarray}
\label{eq:U_pq}
\langle p|U|q\rangle
&=&-\langle p|\omega ^{\dagger}(1+\omega \omega ^{\dagger})^{-1/2}
|q\rangle \nonumber \\
&=&-\sum_{k=1}^{d}(1+\mu_{k}^{2})^{-1/2}\mu _{k}
\langle p|\alpha _{k}\rangle \langle \nu _{k}|q\rangle ,
\end{eqnarray}
and
\begin{eqnarray}
\label{eq:U_q'q}
\langle q'|U|q\rangle
&=&\langle q'|(1+\omega \omega ^{\dagger})^{-1/2}|q\rangle \nonumber \\
&=&\sum_{k=1}^{d}\{(1+\mu_{k}^{2})^{-1/2}-1\}
\langle q'|\nu _{k}\rangle \langle \nu _{k}|q\rangle \nonumber \\
&&+ \delta _{q,q'}.
\end{eqnarray}
Thus, the matrix elements of the effective interaction $\tilde{v}_{12}$
in Eq.~(\ref{eq:eff_int}) can be written as
\begin{equation}
\label{eq:eff_mat}
\langle i|\tilde{v}_{12}|j \rangle = \sum_{k,l}\langle i|U^{-1}|k\rangle
\langle k|h_{0}+v_{12}|l\rangle \langle l|U|j\rangle -\langle i|h_{0}|j\rangle,
\end{equation}
where $|i\rangle$, $|j\rangle$, $|k\rangle$,
and $|l\rangle$ denote the basis states in the $P+Q$ space.

The above formulation is employed for deriving the effective interaction
in the present study.
Here we note that since we treat a many-body system, the single-particle
potential $u_{1}$ ($u_{2}$) in $h_{0}$ for both particle and hole states
is introduced to obtain a good unperturbed energy.
In the following, a procedure for determining the effective interaction and
the single-particle potential is given.

\subsection{Two-step method for the calculation of effective interaction}

In nuclear many-body problems, how to determine the single-particle
potential for particle (unoccupied) states as well as hole (occupied)
states is important
in connection with the evaluation of many-body
correlations~\cite{Jeukenne76,Kuo86,Suzuki87_PTP,Song98}.
In our calculations, the single-particle potential,
which is determined self-consistently
with the two-body effective interaction,
is calculated up to a sufficiently high-momentum region.
In general, this choice of the single-particle potential leads to a deeper
binding of the ground-state energy of a nucleus in the lowest order.
Then, effects of the many-body correlations of higher order become smaller
than the choice of only the kinetic energy for the particle state.
This trend would be favorable when the evaluation of many-body correction terms
has to be limited in the actual calculation.

In our earlier calculations, the effective interaction was derived
by a three-step procedure with some approximations to take account of
single-particle potentials up to a high-momentum region.
In the present work, however, we adopt a two-step procedure and approximation
methods are refined,
because the performance of the computer has been greatly improved
and some approximations in the previous works are not needed at present.
In the following, we shall give the two-step procedure
for the numerical calculation.

\subsubsection{First-step calculation}

\begin{figure}[t]
\includegraphics[width=.300\textheight]{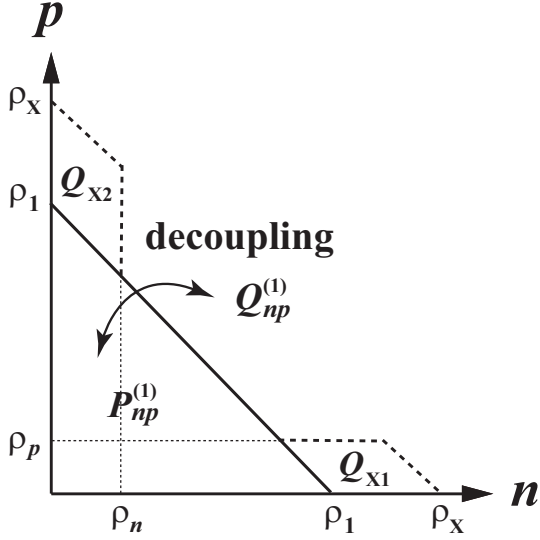}
\caption{\label{fig:model_np} Model space $P_{np}^{(1)}$ and its compliment
$Q_{np}^{(1)}$ for the $np$ channel in the first-step calculation.}
\end{figure}

In this work, we employ the harmonic-oscillator (h.o.) wave functions
as the basis states.
Two-nucleon states for $Z=nn$, $np$, $pp$ channels consisting of the product
of the h.o. states are given by
\begin{equation}
\label{tb_state}
|\alpha \beta \rangle _{Z}=|n_{a}l_{a}j_{a}m_{a},
n_{b}l_{b}j_{b}m_{b}\rangle _{Z},
\end{equation}
The model space $P_{Z}^{(1)}$ and its complement $Q_{Z}^{(1)}$
composed of the two-nucleon states for
the $Z$ channel are defined
with a boundary number $\rho _{1}$ as
\begin{eqnarray}
\label{eq:def_model1}
|\alpha \beta \rangle _{Z}\in
\left\{ \begin{array}{ll}
       P_{Z}^{(1)}
     & {\rm if}\ 2n_{a}+l_{a}+2n_{b}+l_{b}\leq \rho _{1},\\
       Q_{Z}^{(1)}
     & {\rm otherwise},
        \end{array}
\right.
\end{eqnarray}
which is also illustrated only for the $np$
channel in Fig.~\ref{fig:model_np}.
The $nn$ and $pp$ channels are considered similarly in the actual
calculation.
The value of $\rho _{1}$ is taken as large as possible so that
the calculated results do not depend on this value.
The $\rho _{1}$ dependence of calculated results will be investigated
in Sec.~\ref{sec:results}.
The symbols $\rho _{n}$ and $\rho _{p}$ in Fig.~\ref{fig:model_np}
stand for the uppermost occupied states
of the neutron and proton, respectively, and in the present case of $^{16}$O,
$\rho _{n}$ and $\rho _{p}$ are the $0p_{1/2}$ orbits.
The $Q_{\rm X1}$ and $Q_{\rm X2}$ spaces defined with
$\rho _{1}$,
$\rho _{n}$, $\rho _{p}$, and $\rho _{\rm X}$ in the $Q_{np}^{(1)}$ space
should be excluded due to the Pauli principle
when we calculate matrix elements of the bare two-body interaction.
The value of $\rho _{\rm X}$ is determined so that the Pauli principle from
the states in the $Q_{\rm X1}$ and $Q_{\rm X2}$ spaces can well be taken into
account, and taken as $\rho _{\rm X} =2n_{a}+l_{a}+2n_{b}+l_{b}=20$
in the present study.

It is noted that, in this first step, the effective interaction is constructed
using the relative and c.m. states of the h.o. wave functions.
Since we consider a huge Hilbert space, it is very difficult to
use the basis states composed of the product of the single-particle h.o. states
in the model and complementary spaces.
In the following, we shall give a practical method for calculating
the effective interaction and the single-particle potential.
In order to derive the two-body effective interaction for each of
$Z=nn$, $np$, $pp$ channels
we rewrite Eq.~(\ref{eq:eigenvalue_eq})
in terms of the relative and c.m. states as
\begin{eqnarray}
\label{eq:eigenequation_1}
\lefteqn{ [P_{Z}^{(1)}\{ t_{r}+u_{12}(N,L)\}
P_{Z}^{(1)}+Q_{Z}^{(1)}t_{r}Q_{Z}^{(1)}+t_{\rm c.m.}}\nonumber \\
&+&(P_{Z}^{(1)}+\bar{Q}_{Z}^{(1)})v_{12}(P_{Z}^{(1)}+\bar{Q}_{Z}^{(1)})]
|k;l(l')SJ_{r},NL\rangle _{Z}\nonumber \\
&=&E_{k}|k;l(l')SJ_{r},NL\rangle _{Z},
\end{eqnarray}
where $l$ ($l'$) and $S$ are the orbital angular momentum and spin of
a relative state, and $J_{r}$ is the total angular momentum given
by ${\bf J}_{r}={\bf l}+{\bf S}$.
The letter $k$ means an additional quantum number specifying an eigenstate.
The terms $t_{r}$ and $t_{\rm c.m.}$ are the kinetic energies of
the relative and c.m. motions, respectively, and $v_{12}$
is the bare interaction.
The operator $\bar{Q}_{Z}^{(1)}$ projects two-body states on to the
${Q}_{Z}^{(1)}$ space, but Pauli-forbidden two-body states in the
$Q_{X1}$ and $Q_{X2}$ spaces are excluded.
The sum of two single-particle potentials in the relative and c.m. states
is denoted by $u_{12}(N,L)$.
We assume, in the present study, that the matrix elements
in solving Eq.~(\ref{eq:eigenequation_1}) are diagonal
in each of the c.m. quantum numbers $N$ and $L$.
Thus, the resultant effective interaction becomes also diagonal
in the c.m. quantum numbers.

The matrix elements of $u_{12}(N,L)$ can be written under
the angle-average approximation~\cite{Wong67,Suzuki87} as
\begin{eqnarray}
\label{eq:u12_rel}
\lefteqn{\langle nlSJ_{r},NL|u_{12}(N,L)|n'l'SJ_{r},NL\rangle _{Z}}
\nonumber \\
&=&\delta _{l,l'}\sum_{\stackrel
{\scriptstyle n_{a}n_{a}'l_{a}j_{a}n_{b}l_{b}j_{b}}
{\lambda \lambda 'J}}
(-1)^{\lambda +\lambda '}\nonumber \\
&&\times
\left\{ \begin{array}{ccc}
       l_{a}  & \frac{1}{2}  &  j_{a} \\
       l_{b}  & \frac{1}{2}  &  j_{b} \\
     \lambda  &       S      &   J
           \end{array} \right\}
\left\{ \begin{array}{ccc}
       l_{a}  & \frac{1}{2}  &  j_{a} \\
       l_{b}  & \frac{1}{2}  &  j_{b} \\
    \lambda ' &       S      &   J
           \end{array} \right\} \nonumber \\
&&\times\frac{[\lambda][\lambda '][j_{a}][j_{b}][S][J]}
{[L]}W(LlJS;\lambda J_{r})W(Ll'JS;\lambda 'J_{r})\nonumber \\
&&\times\langle nlNL\lambda |n_{a}l_{a}n_{b}l_{b}\lambda \rangle
        \langle n'l'NL\lambda '|n_{a}'l_{a}n_{b}
l_{b}\lambda '\rangle \nonumber \\
&&\times(\langle n_{a}l_{a}j_{a}|u_{z_{1}}^{(1)}|n_{a}'l_{a}j_{a}\rangle
          +\langle n_{a}l_{a}j_{a}|u_{z_{2}}^{(1)}|n_{a}'l_{a}j_{a}\rangle ),
\end{eqnarray}
where $[x]\equiv 2x+1$ and $J$ is the total angular momentum
for two single-particle h.o. states given by
${\bf J}={\bf j_{a}}+{\bf j_{b}}$.
The coefficients $\{ \cdot \cdot \cdot\}$,
$W(\cdot \cdot \cdot)$,
and $\langle nl\cdot \cdot \cdot |n_{a}l_{a}\cdot \cdot \cdot \rangle$
denote the Wigner 9-$j$ symbols, the Racah coefficients,
and the h.o. transformation brackets, respectively.
Note that, as for the $nn$ and $pp$ channels, the calculation should be done
only for $l+S=even$.
The quantities $u_{z_{1}}^{(1)}$ and $u_{z_{2}}^{(1)}$ represent
the single-particle potentials of the neutron $u_{n}^{(1)}$
or proton $u_{p}^{(1)}$ in the first-step calculation, depending on
$Z=nn$, $np$, $pp$ channels.
The single-particle potentials $u^{(1)}_{n}$ and $u^{(1)}_{p}$
are calculated self-consistently with the two-body effective interaction,
which will be shown later.

The operator $\bar{Q}_{Z}^{(1)}$ can be written
under the angle-average approximation as
\begin{eqnarray}
\label{eq:Pauli_op}
\bar{Q}_{Z}^{(1)}&=&\sum_{\stackrel{\scriptstyle nlNLSJ_{r}}
{\rho _{1}<2n+l+2N+L}}\theta _{Z}(n,l,N,L,S,J_{r})\nonumber \\
&&\times|nlSJ_{r},NL\rangle \langle nlSJ_{r},NL|,
\end{eqnarray}
where
\begin{eqnarray}
\label{eq:theta}
\lefteqn{\theta _{Z}(n,l,N,L,S,J_{r})}\nonumber \\
&=&1-\sum_{\stackrel{\scriptstyle ab\lambda \lambda 'J}
{\rho _{1}<2n_{a}+l_{a}+2n_{b}+l_{b}\leq \rho _{\rm X}}}
(-1)^{\lambda +\lambda '}f_{Z}\nonumber \\
&&\times
\left\{ \begin{array}{ccc}
       l_{a}  & \frac{1}{2}  &  j_{a} \\
       l_{b}  & \frac{1}{2}  &  j_{b} \\
     \lambda  &       S      &   J
           \end{array} \right\}
\left\{ \begin{array}{ccc}
       l_{a}  & \frac{1}{2}  &  j_{a} \\
       l_{b}  & \frac{1}{2}  &  j_{b} \\
    \lambda ' &       S      &   J
           \end{array} \right\} \nonumber \\
&&\times\frac{[\lambda][\lambda '][j_{a}][j_{b}][S][J]}{[L]}
W(LlJS;\lambda J_{r})W(LlJS;\lambda 'J_{r})\nonumber \\
&&\times\langle nlNL\lambda |n_{a}l_{a}n_{b}l_{b}\lambda \rangle
        \langle nlNL\lambda '|n_{a}l_{a}n_{b}l_{b}\lambda '\rangle \nonumber \\
\end{eqnarray}
with
\begin{eqnarray}
\label{eq:fz12}
f_{Z}=
\left\{ \begin{array}{ll}
       2
     & {\rm for}\ Z=nn \ {\rm or}\ pp,\\
       1
     & {\rm for}\ Z=np.
        \end{array}
\right.
\end{eqnarray}
Note that, as for the $nn$ and $pp$ channels, the calculation should be done
only for $l+S=even$.
The letter $a$ ($b$) for the summation in Eq.~(\ref{eq:theta})
means a set of the quantum numbers
$a\equiv \{ n_{a},l_{a},j_{a},z=n\ {\rm or}\ p\}$
of a single-particle h.o. state.
The conditions of the summation of single-particle states $a$ and $b$
for the $nn$ and $pp$ channels are
$\{ a\leq \rho _{n}$, $b>\rho_{n}\}$ and
$\{ a\leq \rho _{p}$, $b>\rho_{p}\}$, respectively.
As for the $np$ channel, $\{ a\leq \rho _{n}$, $b>\rho_{p}\}$ or
$\{ a> \rho _{n}$, $b\leq\rho_{p}\}$.
Here for example, the notation $\{ a\leq \rho _{n}$, $b>\rho_{p}\}$
for the $np$ channel means
that the summation is done for occupied states of the neutron and unoccupied
states of the proton.

It should be noted that Eq.~(\ref{eq:eigenequation_1}) is solved exactly
by diagonalizing the matrix elements of several hundred coordinate-space
h.o. basis states for each channel on the assumption of the diagonal c.m.
quantum numbers.
If we employ a bare interaction in momentum-space representation,
the Fourier transformation for the h.o. wave function is needed in calculating
the matrix elements of the bare interaction.
Using the eigenvector $|k;l(l')SJ_{r},NL\rangle _{Z}$, the operator $\omega$
in Eq.~(\ref{eq:omega_qp}) can be written in terms of relative and c.m. states.
Then, the matrices of the effective interaction $\tilde{v}_{12}^{(1)}$
in Eq.~(\ref{eq:eff_mat}) are obtained in the relative and c.m. states as
$\langle nlSJ_{r}|\tilde{v}_{12}(N,L)|n'l'SJ_{r}\rangle _{Z}$
through Eqs.~(\ref{eq:omega_qp})-(\ref{eq:eff_mat}).
Note that we do not need the $Q$-space effective interaction
in the first-step calculation
if we take a sufficiently large model space.

The transformation of the effective interaction in the relative and c.m. states
into the one in the shell-model states can be performed straightforwardly as
\begin{eqnarray}
\label{eq:v12}
\lefteqn{\langle ab|\tilde{v}_{12}^{(1)}|cd\rangle _{J,Z}}\nonumber \\
&=&\frac{1}{\sqrt{1+\delta _{a,b}}}\frac{1}{\sqrt{1+\delta _{c,d}}}
\sum_{\stackrel
{\scriptstyle nln'l'J_{r}S}
{NL\lambda \lambda '}}
(-1)^{\lambda +\lambda '}f_{Z}(l,S)\nonumber \\
&&\times\sqrt{[j_{a}][j_{b}][j_{c}][j_{d}]}
[\lambda][\lambda '][S][J_{r}]\nonumber \\
&&\times\left\{ \begin{array}{ccc}
       l_{a}  & \frac{1}{2}  &  j_{a} \\
       l_{b}  & \frac{1}{2}  &  j_{b} \\
     \lambda  &       S      &   J
           \end{array} \right\}
     \left\{ \begin{array}{ccc}
       l_{c}  & \frac{1}{2}  &  j_{c} \\
       l_{d}  & \frac{1}{2}  &  j_{d} \\
    \lambda ' &       S      &   J
           \end{array} \right\}\nonumber \\
&&\times W(LlJS;\lambda J_{r})W(Ll'JS;\lambda 'J_{r})\nonumber \\
&&\times\langle nlNL\lambda |n_{a}l_{a}n_{b}l_{b}\lambda \rangle
\langle n'l'NL\lambda '|n_{c}l_{c}n_{d}l_{d}\lambda '\rangle \nonumber \\
&&\times\langle nlSJ_{r}|\tilde{v}_{12}^{(1)}(N,L)|n'l'SJ_{r}\rangle _{Z},
\end{eqnarray}
where
\begin{eqnarray}
\label{eq:fzls}
f_{Z}(l,S)=
\left\{ \begin{array}{ll}
       1+(-1)^{l+S}
     & {\rm for}\ Z=nn \ {\rm or}\ pp,\\
       1
     & {\rm for}\ Z=np.
        \end{array}
\right.
\end{eqnarray}
which is required for the antisymmetrization of the matrix elements
in the shell-model states.
Note that $l+S=even$ for the $nn$ and $pp$ channels.

The single-particle potentials $u_{z_{1}}^{(1)}$ and $u_{z_{2}}^{(1)}$
in Eq.~(\ref{eq:u12_rel}) are determined self-consistently with the two-body
effective interaction $v_{12}^{(1)}$, which is written as
\begin{eqnarray}
\label{eq:u_n_1st}
\langle a|u_{n}^{(1)}|a'\rangle
&=&\sum_{\stackrel{\scriptstyle J,Z=nn, np}{m:{\rm occupied}}}
\frac{1}{\sqrt{1+\delta _{a,m}}}\frac{1}
{\sqrt{1+\delta _{a',m}}}(2J+1)\nonumber \\
&&\times\langle am|\tilde{v}_{12}^{(1)}|a'm\rangle _{J,Z}
\end{eqnarray}
for the neutron, and
\begin{eqnarray}
\label{eq:u_p_1st}
\langle a|u_{p}^{(1)}|a'\rangle
&=&\sum_{\stackrel{\scriptstyle J,Z=pp, np}{m:{\rm occupied}}}
\frac{1}{\sqrt{1+\delta _{a,m}}}\frac{1}
{\sqrt{1+\delta _{a',m}}}(2J+1)\nonumber \\
&&\times \langle ma|\tilde{v}_{12}^{(1)}|ma'\rangle _{J,Z}
\end{eqnarray}
for the proton.

The procedure for the self-consistent calculation is as follows.
First, we input initial values of $u_{z_{1}}^{(1)}$ and $u_{z_{2}}^{(1)}$
in Eq.~(\ref{eq:u12_rel}),
and solve the eigenvalue equation in Eq.~(\ref{eq:eigenequation_1}) for each of
$Z=nn$, $np$, $pp$ channels.
Through Eqs.~(\ref{eq:omega_qp})-(\ref{eq:eff_mat}), the effective interaction
in the form of the reduced matrix element is determined.
Then, the new single-particle potentials are calculated
through Eqs.~(\ref{eq:v12})-(\ref{eq:u_p_1st}).
These new values of the single-particle potentials
are used in Eq.~(\ref{eq:u12_rel}), and
the iterative calculation is performed until the calculated results converge.

We remark here that one of the practical methods of the structure calculations
using the present effective interaction
would be the shell-model diagonalization.
However, the application of such a calculation may be limited
only to light nuclei,
because we must take account of many single-particle states in the model space
and the dimension of the matrices to be diagonalized becomes very huge.
Since we intend to obtain only the energies of the ground state of
the closed-shell nucleus and the single-particle (-hole) states
in its neighboring nuclei,
we proceed to the next step for a more practical calculation.
In the second-step calculation, the effective interaction determined
in the first-step calculation
is unitarily transformed again so as to vanish the matrix elements
for two-particle two-hole ($2p2h$) excitation.
This is an essential point of the UMOA.
By virtue of this, a number of many-body correlations with the vertices of
the effective interaction are reduced compared to the usual linked-cluster
expansion with the $G$ matrix.
In the UMOA, such many-body correlations can be evaluated
in a cluster expansion of the unitarily transformed Hamiltonian with
the vertices of $S$ in Eq.~(\ref{S}), the one-body Hamiltonian,
and the two-body effective interaction.

\subsubsection{Second-step calculation}

Using the two-body effective interaction $\tilde{v}_{ij}^{(1)}$
determined in the first-step calculation,
we consider the internal Hamiltonian as
\begin{equation}
\label{eq:H_int}
\tilde{H}_{\rm int}=\sum_{i}t_{i}+\sum_{i<j}\tilde{v}^{(1)}_{ij}-T_{\rm c.m.},
\end{equation}
where $T_{\rm c.m.}$ is the kinetic energy of the c.m. motion.
In this second step, the calculations are performed
using the basis states of the product of the single-particle h.o. states.
In order to remove spurious c.m. states,
we add the c.m. Hamiltonian $H_{\rm c.m.}$ so as to constrain
the ground-state c.m. motion
in the h.o. potential with the frequency $\Omega$ as
\begin{equation}
\label{eq:H_cm}
H_{\rm c.m.}=\beta _{\rm c.m.}\left( T_{\rm c.m.}
+U_{\rm c.m.}-\frac{3}{2}\hbar\Omega \right).
\end{equation}
The h.o. potential $U_{\rm c.m.}$ can be written with the mass number $A$
and the nucleon mass $m$ as
\begin{eqnarray}
\label{eq:hopot_cm}
U_{\rm c.m.}&=&\frac{1}{2}Am\Omega ^{2}{\bf R}^{2}\nonumber \\
&=&\sum_{i<j}\left( \frac{1}{A-1}X_{ij}-\frac{A-2}{A(A-1)}x_{ij}\right),
\end{eqnarray}
where $X_{ij}=\frac{1}{2}(2m)\Omega ^{2}{\bf R}_{ij}^{2}$ and
$x_{ij}=\frac{1}{2}(\frac{m}{2})\Omega ^{2}{\bf r}_{ij}^{2}$.
The definitions of the coordinates are
${\bf R}=\frac{1}{A}\sum_{i}{\bf r}_{i}$,
${\bf R}_{ij}=\frac{1}{2}({\bf r}_{i}+{\bf r}_{j})$,
and ${\bf r}_{ij}={\bf r}_{i}-{\bf r}_{j}$.
We assume that the nucleon mass is the mean value of the neutron and proton.
As for the value of $\beta _{\rm c.m.}$ in Eq.~(\ref{eq:H_cm}),
in the present study, we simply take as
$\beta _{\rm c.m.}=1$ which could be acceptable as discussed
in Refs.~\cite{Dean99,Mihaila00}.
Thus, the Hamiltonian to be considered
in the second-step calculation becomes
\begin{eqnarray}
\label{eq:H_2nd}
\tilde{H}&=&H_{\rm int}+H_{\rm c.m.}\nonumber \\
&=&\sum_{i}t_{i}+\sum_{i<j}\tilde{V}_{ij}^{(1)}(A)-\frac{3}{2}\hbar\Omega ,
\end{eqnarray}
where
\begin{equation}
\label{eq:V_ij}
\tilde{V}_{ij}^{(1)}(A)=\tilde{v}_{ij}^{(1)}+\frac{1}{A-1}X_{ij}
-\frac{A-2}{A(A-1)}x_{ij}.
\end{equation}
Note that the above two-body interaction is $A$ dependent.

\begin{figure}[t]
\includegraphics[width=.330\textheight]{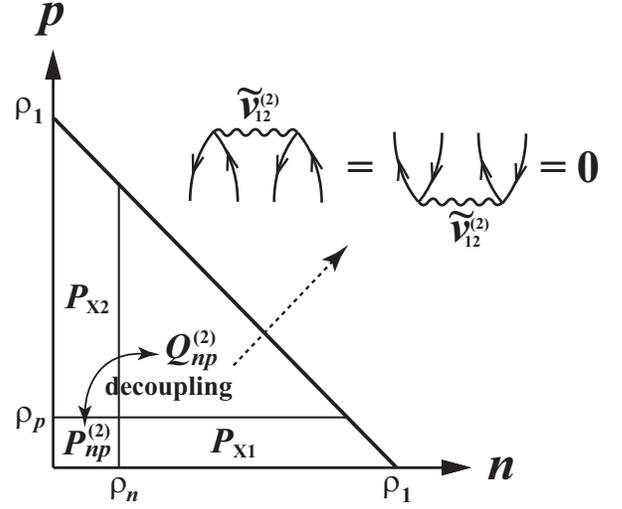}
\caption{\label{fig:model_np2} Models space $P_{np}^{(2)}$ and its compliment
$Q_{np}^{(2)}$ for the $np$ channel in the second-step calculation.}
\end{figure}

The central aim of the present study is to calculate the binding energies of
the ground-state of $^{16}$O and its neighboring nuclei, and to obtain
single-particle energies using the Hamiltonian in Eq.~(\ref{eq:H_2nd}).
To accomplish this without performing the full shell-model diagonalization,
we proceed to the decoupling calculation again.
The model space in the first-step calculation in Fig.~\ref{fig:model_np}
is separated as shown in Fig.~\ref{fig:model_np2}.
The $model$ space and its $compliment$ for the $np$ channel are denoted by
$P_{np}^{(2)}$ and $Q_{np}^{(2)}$, respectively.
It should be noted that we solve the $P$-space and $Q$-space problems
on an equal footing in the second-step calculation,
using the effective interaction determined in the first-step calculation
which has already incorporated the effect of the short-range correlation of
the bare interaction.
The $P_{X1}$ and $P_{X2}$ spaces are the Pauli-blocked spaces
in the second-step calculation.

It would be worthy to mention the property of the effective interaction
to be determined in the second-step calculation.
By taking the model and complementary spaces as shown
in Fig.~\ref{fig:model_np2},
the resultant effective interaction $\tilde{v}_{12}^{(2)}$
which is determined from
the decoupling condition $Q_{Z}^{(2)}\tilde{v}_{12}^{(2)}P_{Z}^{(2)}=0$
for $Z=nn$, $np$, $pp$ has no vertices which induce $2p2h$ excitation.
This is analogous to the Hartree-Fock (HF) condition which means
that an original Hamiltonian is transformed
so as to vanish the vertices of $1p1h$ excitation.
Although the vertices of the one-body non-diagonal matrix elements remain
in determining the effective interaction,
these non-diagonal matrix elements are diagonalized
at the end of the calculation.

The eigenvalue equation for the $Z$ channel in the second-step calculation
which corresponds to Eq.~(\ref{eq:eigenvalue_eq}) can be
written as
\begin{equation}
\label{eq:eigenequation_2}
\{ t_{z_{1}}+u_{z_{1}}^{(2)}+t_{z_{2}}+u_{z_{2}}^{(2)}
+\tilde{V}_{12}^{(1)}(A)\}
|\Psi _{k}\rangle _{J^{\pi},Z}=E_{k}|\Psi _{k}\rangle _{J^{\pi},Z},
\end{equation}
where $|\Psi _{k}\rangle _{J^{\pi},Z}$ represents a two-body eigenstate
in terms of the basis states of the product of the single-particle h.o. states
with a good total angular momentum and parity for the $Z$ channel.
We solve the above eigenvalue equation exactly by diagonalizing
the matrix elements in the $full$ space $P_{Z}^{(2)}+Q_{Z}^{(2)}$,
and then obtain the matrix elements of $U$ in this $full$ space through
Eqs.~(\ref{eq:omega_qp})-(\ref{eq:U_q'q}).
In addition, the matrix elements of $U$ for the $P_{X1}$ and $P_{X2}$ spaces
are given by
\begin{equation}
\label{eq:U_x'x}
\langle x'|U|x\rangle =\delta _{x,x'}
\end{equation}
and
\begin{equation}
\label{eq:U_0}
\langle p|U|x\rangle =\langle q|U|x\rangle
=\langle x|U|p\rangle =\langle x|U|q\rangle =0,
\end{equation}
where $|x\rangle$, $|p\rangle$, and $|q\rangle$ are the basis states in the
$P_{X1}$ and $P_{X2}$, $P_{Z}^{(2)}$, and $Q_{Z}^{(2)}$ spaces, respectively.

The calculation procedure in the second step is as follows.
We first solve exactly Eq.~(\ref{eq:eigenequation_2}) by the diagonalization.
As the initial values of $u_{z_{1}}^{(2)}$ and $u_{z_{2}}^{(2)}$
in Eq.~(\ref{eq:eigenequation_2}), we use the single-particle potentials
determined in the first-step calculation.
Through Eqs.~(\ref{eq:omega_qp})-(\ref{eq:eff_mat}), the effective interaction
$\tilde{v}_{12}^{(2)}$ in this second step in all the
$P_{X1}$, $P_{X2}$, $P_{Z}^{(2)}$, and $Q_{Z}^{(2)}$ spaces is determined.
Then, the single-particle potentials $u_{n}^{(2)}$ and $u_{p}^{(2)}$
are calculated in Eqs.~(\ref{eq:u_n_1st})-(\ref{eq:u_p_1st})
using the effective interaction $\tilde{v}_{12}^{(2)}$
instead of $\tilde{v}_{12}^{(1)}$ determined in the first step,
and the self-consistent calculation is performed iteratively
until the calculated results converge.

\begin{figure}[t]
\includegraphics[width=.300\textheight]{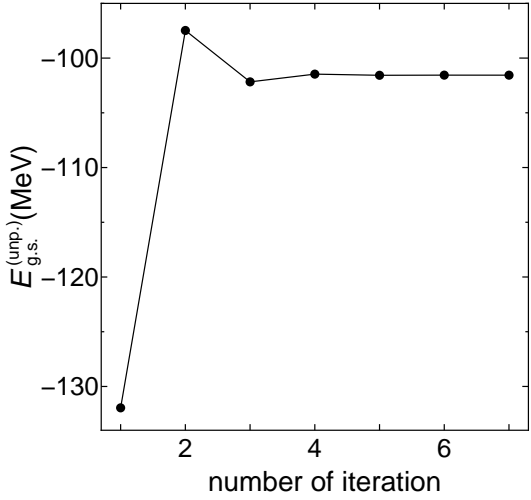}
\caption{\label{fig:3} Convergence of the unperturbed ground-state energy for
the iterative calculation for $^{16}$O
in the second-step calculation for $\rho _{1}=12$
and $\hbar \Omega=14$ MeV. The CD-Bonn potential is employed.}
\end{figure}

As a typical example of the convergence of the self-consistent calculation,
in Fig.~\ref{fig:3}, we show the results of the unperturbed ground-state
energy of $^{16}$O in the second-step calculation with increasing
number of iteration for $\rho _{1}=12$ with the CD-Bonn potential.
The Coulomb interaction is included in the calculation.
The unperturbed ground-state energy $E_{\rm g.s.}^{\rm (unp.)}$
of the doubly closed-shell nucleus is given by
\begin{eqnarray}
\label{eq:gs_energy}
\lefteqn{E_{\rm g.s.}^{\rm (unp.)}}\nonumber \\
&=&\sum_{\stackrel{\scriptstyle z=n,p}{m:{\rm occupied}}}
(2j_{m}+1)\left( \langle m|t_{z}|m\rangle
+\frac{1}{2}\langle m|u_{z}^{(2)}|m\rangle\right)\nonumber \\
&&-\frac{3}{2}\hbar \Omega ,
\end{eqnarray}
where $|m\rangle$ denotes a h.o. single-particle state for
the hole state
with a total angular momentum $j_{m}$.
We also express the unperturbed single-particle energy
$E_{\rm s.p.}^{\rm (unp.)}$ for a state $|a\rangle$ as
\begin{equation}
\label{eq:SPE_UNP}
E_{\rm s.p.}^{\rm (unp.)}=\langle a|t_{z}|a\rangle
+\langle a|u_{z}^{(2)}|a\rangle\ {\rm for\ } z=n,\ p.
\end{equation}
Note that $E_{\rm g.s.}^{\rm (unp.)}$ and $E_{\rm s.p.}^{\rm (unp.)}$
are implicitly $A$ dependent
due to the property of $\tilde{V}_{ij}^{(1)}(A)$ in Eq.~(\ref{eq:V_ij}).
We see that the results in Fig.~\ref{fig:3} converge
when the number of interaction is larger than 4.

\subsection{Diagonalization of the transformed Hamiltonian}
The transformed Hamiltonian determined in the second-step calculation
does not contain the interaction which induces $2p2h$ excitation.
However, there remain some terms inducing $1p1h$ excitation in the
one-body Hamiltonian, and coupling terms in the two-body interaction
between $1h$ and $1p2h$ states for occupied states,
and between $1p$ and $2p1h$ states for unoccupied states.
The transformed Hamiltonian to be diagonalized consists of the kinetic
and single-particle potential parts,
and the two-body effective interaction determined
in the second-step calculation.
As for the closed-shell nucleus,
we diagonalize the transformed Hamiltonian with the shell-model basis states,
taking into account $1p1h$ excitation from the unperturbed ground state.
We denote the energy shift from the unperturbed energy obtained by
the diagonalization by $E_{1p1h}$.
As for the closed-shell nucleus plus one-particle (one-hole) system,
the shell-model basis states are composed of the $1p$ and $2p1h$ states
($1h$ and $1p2h$ states). 
The energy shift from the unperturbed energy obtained by the
diagonalization is expressed by $E_{2p1h}$ ($E_{1p2h}$).
The binding energies $BE$ for these systems are given as follows.
\begin{equation}
\label{eq:BE_O16}
-BE(^{16}{\rm O})=E_{\rm g.s.}^{\rm (unp.)}+E_{\rm 1p1h},
\end{equation}
\begin{equation}
\label{eq:BE_O17}
-BE({\rm ^{17}O}, {\rm ^{17}F})=E_{\rm g.s.}^{\rm (unp.)}+E_{\rm 2p1h},
\end{equation}
and
\begin{equation}
\label{eq:BE_O15}
-BE({\rm ^{15}O}, {\rm ^{15}N})=E_{\rm g.s.}^{\rm (unp.)}+E_{\rm 1p2h}.
\end{equation}
Thus, the single-particle energies $E_{\rm s.p.}$ for the particle
and hole states are written, respectively, as
\begin{equation}
\label{eq:SPE_P}
E_{\rm s.p.}({\rm ^{17}O}, {\rm ^{17}F})
=BE(^{16}{\rm O})-BE({\rm ^{17}O}, {\rm ^{17}F})
\end{equation}
and
\begin{equation}
\label{eq:SPE_H}
E_{\rm s.p.}({\rm ^{15}O}, {\rm ^{15}N})
=BE({\rm ^{15}O}, {\rm ^{15}N})-BE(^{16}{\rm O}).
\end{equation}

In the following section, we shall present the calculated results
of the energies using Eqs.~(\ref{eq:gs_energy})-(\ref{eq:SPE_H})
with some discussions.

\section{\label{sec:results}Results and discussion}

In the present study,
the number of the h.o. wave functions which are used as the basis states
is finite,
and some approximations are made.
Therefore, the calculated results have the dependences on the h.o. energy
$\hbar \Omega$ and the value of $\rho _{1}$ which specifies the model
space in the first-step calculation.
In the following subsections, some calculated results are shown with the
$\hbar \Omega$ and $\rho _{1}$ dependences.
However, we search for optimal values of $\hbar \Omega$ and
values of $\rho _{1}$ for which the calculated results almost converge
to obtain the final results.

In order to clarify differences in the properties of modern nucleon-nucleon
interactions, four interactions represented in momentum space
are employed, namely, the Nijm-93,
Nijm-I~\cite{Stoks94}, the CD-Bonn~\cite{Machleidt96} and
the N$^{3}$LO~\cite{Entem03} potentials,
and the Coulomb force is also used commonly.
In the calculations, the partial waves up to $J_{r}\leq 6$
are taken into account.

\subsection{\label{sec:O16}$^{16}$O}

\begin{figure}[b]
\includegraphics[width=.365\textheight]{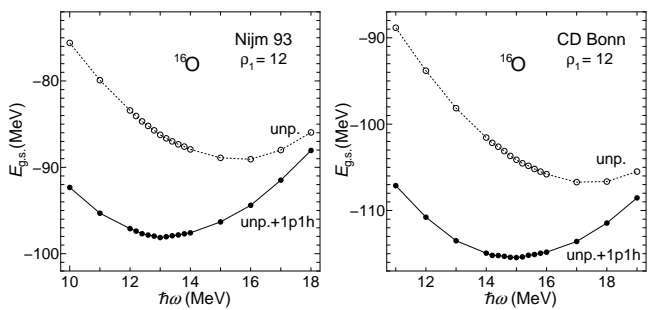}
\caption{\label{fig:O16_hw} The $\hbar \Omega$ dependence of calculated
ground-state energies
of $^{16}$O for $\rho _{1}=12$ for the Nijm-93 and the CD-Bonn potentials.}
\end{figure}

In Fig.~\ref{fig:O16_hw}, the $\hbar \Omega$ dependence of calculated
ground-state energies of $^{16}$O for $\rho _{1}=12$ using the Nijm-93
and the CD-Bonn potentials is shown. 
The unperturbed energy which is shown as ``unp." and the energy
with the $1p1h$ correction are displayed separately.
The expression of the ground-state energy with the $1p1h$ effect
is given in Eq.~(\ref{eq:BE_O16}) as $-BE$.
We see that the effect of the $1p1h$ correction has a significant
contribution attractively to the ground-state energy.
If we use the HF wave functions, the unperturbed ground-state energies
should become more attractive.

We note here that the values of $\hbar \Omega$ at which the energy minima
are obtained differ from each other between the Nijm-93
and the CD-Bonn potentials,
and also between the unperturbed
and the unperturbed plus $1p1h$ energies, reflecting differences in
the properties of the two potentials.
In the calculation of $^{16}$O,
a value around $\hbar \Omega =14$ MeV is often employed as a suitable value
of $\hbar \Omega$.
This value is very close to that determined by empirical formula such as
$\hbar \Omega = 45A^{-1/3}-25A^{-2/3}$ MeV.
In the present study, however, we regard the value at which the energy minimum
is obtained as the optimal one.
The optimal value should be searched for each state in nuclei.

\begin{figure}[t]
\includegraphics[width=.365\textheight]{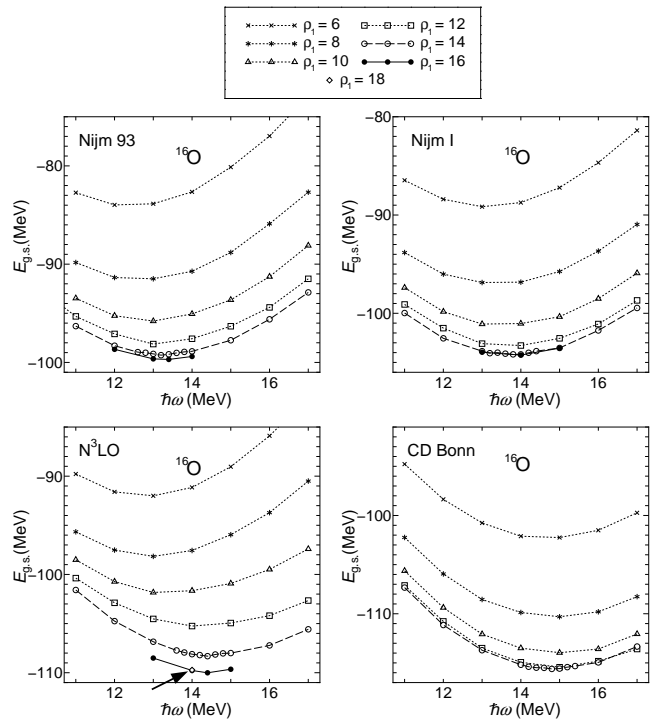}
\caption{\label{fig:O16_hw_rho1} The $\hbar \Omega$ and $\rho _{1}$ dependences
of calculated ground-state energies with the $1p1h$ effect of $^{16}$O
for various modern nucleon-nucleon interactions.}
\end{figure}

Figure~\ref{fig:O16_hw_rho1} illustrates the $\hbar \Omega$ and $\rho _{1}$
dependences of the ground-state energy with the $1p1h$ effect
for the Nijm-93, Nijm-I, the N$^{3}$LO, and the CD-Bonn potentials.
In principle, we should take the value of $\rho _{1}$ as large as possible
until the results do not depend on $\rho _{1}$.
When we take as $\rho _{1}=14$, the results show fairly good convergence for
the CD-Bonn potential.
As for the Nijm-93, Nijm-I and the N$^{3}$LO potentials,
almost convergent results are obtained
if we take the value $\rho _{1}=16$.
Note that the energy for $\rho _{1}=18$ at $\hbar \Omega=14$ MeV is calculated
for the N$^{3}$LO
potential in order to confirm the convergence.
We see that the results for $\rho _{1}=16$ and 18 are almost the same.

In Table \ref{tab:O16_gs}, the final results of the ground-state energy
with the $1p1h$ effect are tabulated for the four potentials
together with the experimental value.
The binding energies per nucleon are also shown.
The calculated values are for the optimal $\hbar \Omega$ and $\rho _{1}$
which can be
determined from the results as shown in Fig.~\ref{fig:O16_hw_rho1}.
The results for the Nijm~93 and the CD~Bonn are the least and most attractive,
respectively, of the four potentials.
This tendency can also be observed in the Faddeev-Yakubovsky calculations
for $^{4}$He by Nogga $et$ $al$~\cite{Nogga02}.

\begin{table}[t]
\caption{\label{tab:O16_gs} The calculated ground-state
energies with the $1p1h$ effect and the binding energies per nucleon
of $^{16}$O.
In these calculated values, the optimal values of $\hbar \Omega$
for each interaction
are employed.
As for the value of $\rho _{1}$, we take as $\rho _{1}=14$ for the CD~Bonn
and $\rho _{1}=16$ for the other interactions as the sufficiently large value
as suggested in Fig.~\ref{fig:O16_hw_rho1}.
The experimental values are taken from Ref.~\cite{Audi93}.
All energies are in MeV.}
\begin{ruledtabular}
    \begin{tabular}{cccccc}
      $^{16}$O        &   Nijm 93  &    Nijm I   &    N$^{3}$LO   &   CD Bonn   &    Expt.   \\ \hline
      $ E_{\rm g.s.} $ & $ -99.69 $ & $ -104.25 $ & $ -110.00 $ & $ -115.61 $ & $ -127.62 $ \\
      $ BE/A $        & $   6.23 $ & $    6.52 $ & $    6.88 $ & $    7.23 $ & $    7.98 $ \\
    \end{tabular}
\end{ruledtabular}
\end{table}

It is seen that the calculated ground-state energies are less bound
than the experimental value.
In the present calculation, higher-order correlations
such as the three-body cluster terms have not been evaluated.
In addition, the real three-body force is not taken into account.
The inclusion of the real three-body force and the higher-order
many-body correlations would compensate for the discrepancies
between the experimental and calculated values.
Such a study remains as an important task for a deeper understanding of
nuclear ground-state properties in the present approach.
A coupled-cluster calculation of the saturation property concerning
the binding energy and charge radius for $^{16}$O by Mihaila and Heisenberg
has shown that the calculated result agrees well with the experimental value
when a genuine three-body force is included in the calculation~\cite{Mihaila00}.

\subsection{\label{sec:N15_O15}$^{15}$N and $^{15}$O}

\begin{figure}[b]
\includegraphics[width=.365\textheight]{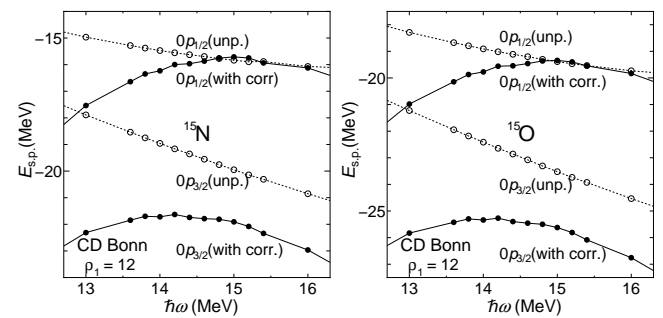}
\caption{\label{fig:N15_O15_hw} The $\hbar \Omega$ dependence of calculated
single-particle energies for $\rho _{1}=12$ for hole states in $^{16}$O.
The left (right) figure is for the proton (neutron) levels which correspond to
the single-hole states in $^{15}$N ($^{15}$O).
The CD-Bonn potential is employed.
}
\end{figure}

Shown in Fig.~\ref{fig:N15_O15_hw} is the $\hbar \Omega$
dependence of calculated single-particle energies for the 0$p$ states
in $^{15}$N and $^{15}$O for $\rho _{1}=12$
in the case of the CD-Bonn potential.
The unperturbed energy and the energy with the $1p2h$ correction
are displayed separately.
The unperturbed single-particle energy is given in Eq.~(\ref{eq:SPE_UNP})
and that with the $1p2h$ correction is in Eq.~(\ref{eq:SPE_H}).
We see that the unperturbed single-particle energies vary considerably
at around the typical $\hbar \Omega =14$ MeV.
However, the single-particle energies with the $1p2h$ correction have the
saturation points at around $\hbar \Omega\simeq 14$ $\sim$ $15$ MeV,
depending on the single-particle states.
Note that the minimum points for the ground-state energies
of $^{15}$N and $^{15}$O
correspond to the maximum points of the single-particle energies
in Fig.~\ref{fig:N15_O15_hw}.
It should be remarked that the spin-orbit splittings for the 0$p$ states
in $^{15}$N and $^{15}$O are significantly enlarged by taking into account
the $1p2h$ correction.
This effect has already shown in our previous works though the calculation
was performed perturbatively by taking into account
second-order diagrams on the isospin basis~\cite{Suzuki94,Suzuki87}.

\begin{figure}[t]
\includegraphics[width=.365\textheight]{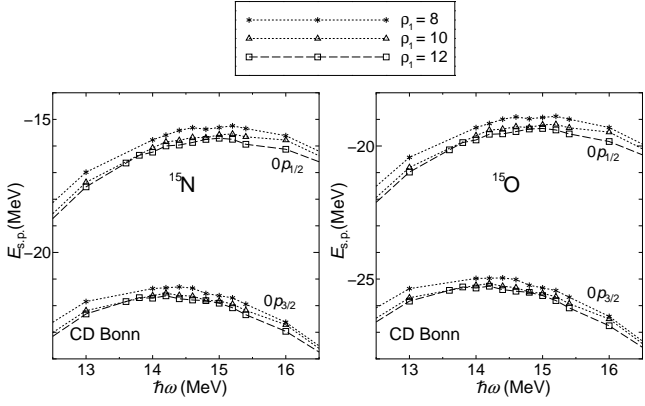}
\caption{\label{fig:N15_O15_hw_rho1} The $\hbar \Omega$ and $\rho _{1}$
dependences of calculated single-particle energies with the $1p2h$ effect
for the proton ($^{15}$N) and neutron ($^{15}$O) levels.
The CD-Bonn potential is employed.
}
\end{figure}

Figure~\ref{fig:N15_O15_hw_rho1} exhibits the $\hbar \Omega$ and $\rho _{1}$
dependences of the single-particle energies with the $1p2h$ effect.
It is seen that the $\rho _{1}$ dependence is weaker than
that in Fig.~\ref{fig:O16_hw_rho1}.
This is because the results in Fig.~\ref{fig:N15_O15_hw}
are the relative values
for the binding energies of two nuclei as given in Eq.~(\ref{eq:SPE_H}),
while those in Fig.~\ref{fig:O16_hw_rho1} are not the relative ones.
We may say that, as for the single-particle energies for the hole states,
the results for
$\rho _{1}=12$ are acceptable as final results in the present study.

\begin{figure}[t]
\includegraphics[width=.330\textheight]{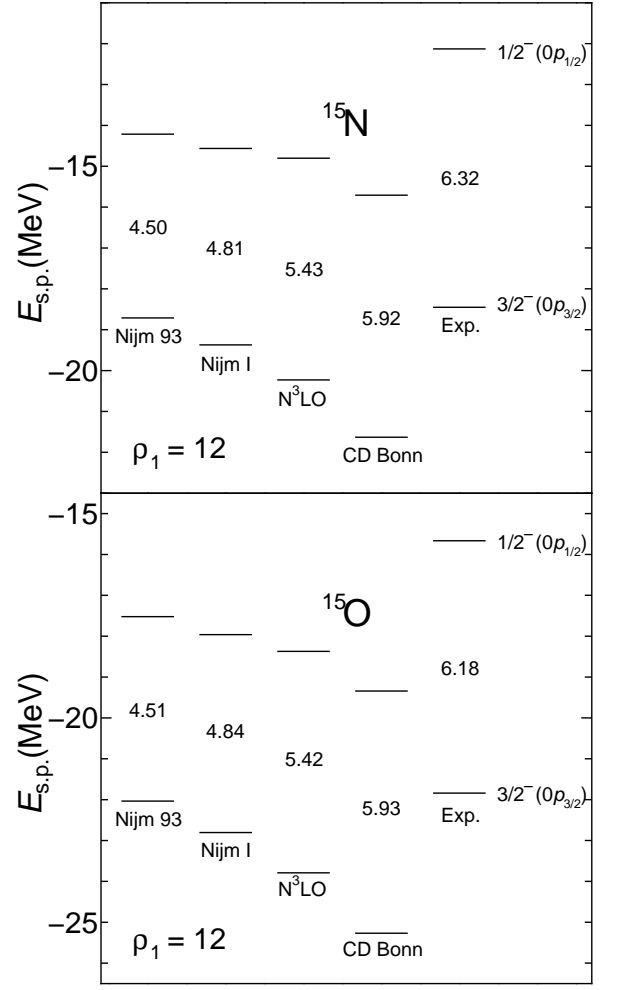}
\caption{\label{fig:N15_O15_sp} The calculated single-particle energies
with the $1p2h$ effect for $\rho _{1}=12$
in $^{15}$N and $^{15}$O.
The values of the spin-orbit splitting are also shown.
In these calculated values, the optimal values of $\hbar \Omega$
for each single-hole state and interaction are employed.
}
\end{figure}

In Fig.~\ref{fig:N15_O15_sp}, the final results of the single-particle energies
with the $1p2h$ effect for the 0$p$ states in $^{15}$N and $^{15}$O for
$\rho _{1}=12$ using the four potentials are shown with the values of
the spin-orbit splitting energy.
The optimal values of $\hbar \Omega$ for each interaction are searched
for the binding energies of $^{16}$O, $^{15}$N, and $^{15}$O in calculating
the single-particle energies through Eq.~(\ref{eq:SPE_H}).
We see that the calculated spin-orbit splittings are smaller than
the experimental values though the differences between the calculated and
experimental values depend on the nucleon-nucleon interactions employed.
The magnitudes of these discrepancies would be reduced
if we include a genuine three-body force
in the calculation as discussed in Refs.~\cite{Ando81,Pieper93}.

In Tables~\ref{tab:N15_sp} and \ref{tab:O15_sp}, the final results
of the single-particle energies shown
in Fig.~\ref{fig:N15_O15_sp} are tabulated.
The results for the typical $\hbar \Omega=14$ MeV are also displayed
in parentheses for reference.
In the case of $\hbar \Omega=14$ MeV, we use this value commonly
in calculating
the binding energies of $^{16}$O, $^{15}$N, and $^{15}$O.
It is seen that all the calculated single-particle energies
are more attractive than the experimental values.
The inclusion of the three-body force and the evaluation of higher-order
many-body correlations may compensate
for the discrepancies between the experimental and calculated values.

\begin{table}[t]
\caption{\label{tab:N15_sp} The calculated single-particle energies with the $1p2h$ effect
for $\rho _{1}=12$ in $^{15}$N.
The values of the spin-orbit splitting energy
$\Delta E_{ls}(0p)=E_{\rm s.p.}(0p_{1/2})-E_{\rm s.p.}(0p_{3/2})$ are also tabulated.
In these calculated values, the optimal values of $\hbar \Omega$
for each single-hole state and interaction are employed.
The results for $\hbar \Omega =14$ MeV are also shown in parentheses.
The experimental values are taken from Ref.~\cite{AJZENBERG91}.
All energies are in MeV.}
\begin{ruledtabular}
    \begin{tabular}{cccccc}
 $^{15}$N              &   Nijm 93  &   Nijm I   &    N$^{3}$LO  &   CD Bonn   &      Expt.   \\ \hline
 $ 1/2^{-}(0p_{1/2}) $ & $ -14.21 $ & $ -14.56 $ & $ -14.80 $ & $ -15.71 $ & $ -12.13 $ \\
                       & $(-14.12)$ & $(-14.56)$ & $(-14.80)$ & $(-15.73)$ &            \\
 $ 3/2^{-}(0p_{3/2}) $ & $ -18.71 $ & $ -19.37 $ & $ -20.23 $ & $ -21.63 $ & $ -18.45 $ \\
                       & $(-19.26)$ & $(-19.86)$ & $(-20.17)$ & $(-21.22)$ &            \\
 $ \Delta E_{ls}(0p) $ & $   4.50 $ & $   4.81 $ & $   5.43 $ & $   5.92 $ & $   6.32 $ \\
                       & $ ( 5.14)$ & $ ( 5.30) $ & $ (5.37) $ & $ (5.49) $ &            \\
    \end{tabular}
\end{ruledtabular}
\end{table}

\begin{table}[t]
\caption{\label{tab:O15_sp} Same as Table \ref{tab:N15_sp}, except for $^{15}$O.
}
\begin{ruledtabular}
    \begin{tabular}{cccccc}
 $^{15}$O              &   Nijm 93  &   Nijm I   &    N$^{3}$LO   &   CD Bonn  &     Expt.   \\ \hline
 $ 1/2^{-}(0p_{1/2}) $ & $ -17.52 $ & $ -17.96 $ & $ -18.37 $ & $ -19.34 $ & $ -15.66 $ \\
                       & $(-17.51)$ & $(-18.00)$ & $(-18.34)$ & $(-19.27)$ &            \\
 $ 3/2^{-}(0p_{3/2}) $ & $ -22.03 $ & $ -22.80 $ & $ -23.79 $ & $ -25.27 $ & $ -21.84 $ \\
                       & $(-22.72)$ & $(-23.37)$ & $(-23.79)$ & $(-24.83)$ &            \\
 $ \Delta E_{ls}(0p) $ & $   4.51 $ & $   4.84 $ & $   5.42 $ & $   5.93 $ & $   6.18 $ \\
                       & $  (5.21) $ & $ (5.37) $ & $ (5.45) $ & $ (5.56)$ &            \\
    \end{tabular}
\end{ruledtabular}
\end{table}

In order to see the accuracy of the calculations,
it would be worthwhile to apply the present method to the few-nucleon systems
$^{4}$He, $^{3}$H and $^{3}$He as similar systems to
$^{16}$O, $^{15}$N, and $^{15}$O.
As for the few-nucleon systems, the binding energies have been calculated
precisely by various methods~\cite{Kamada01}.

\begin{figure}[t]
\includegraphics[width=.365\textheight]{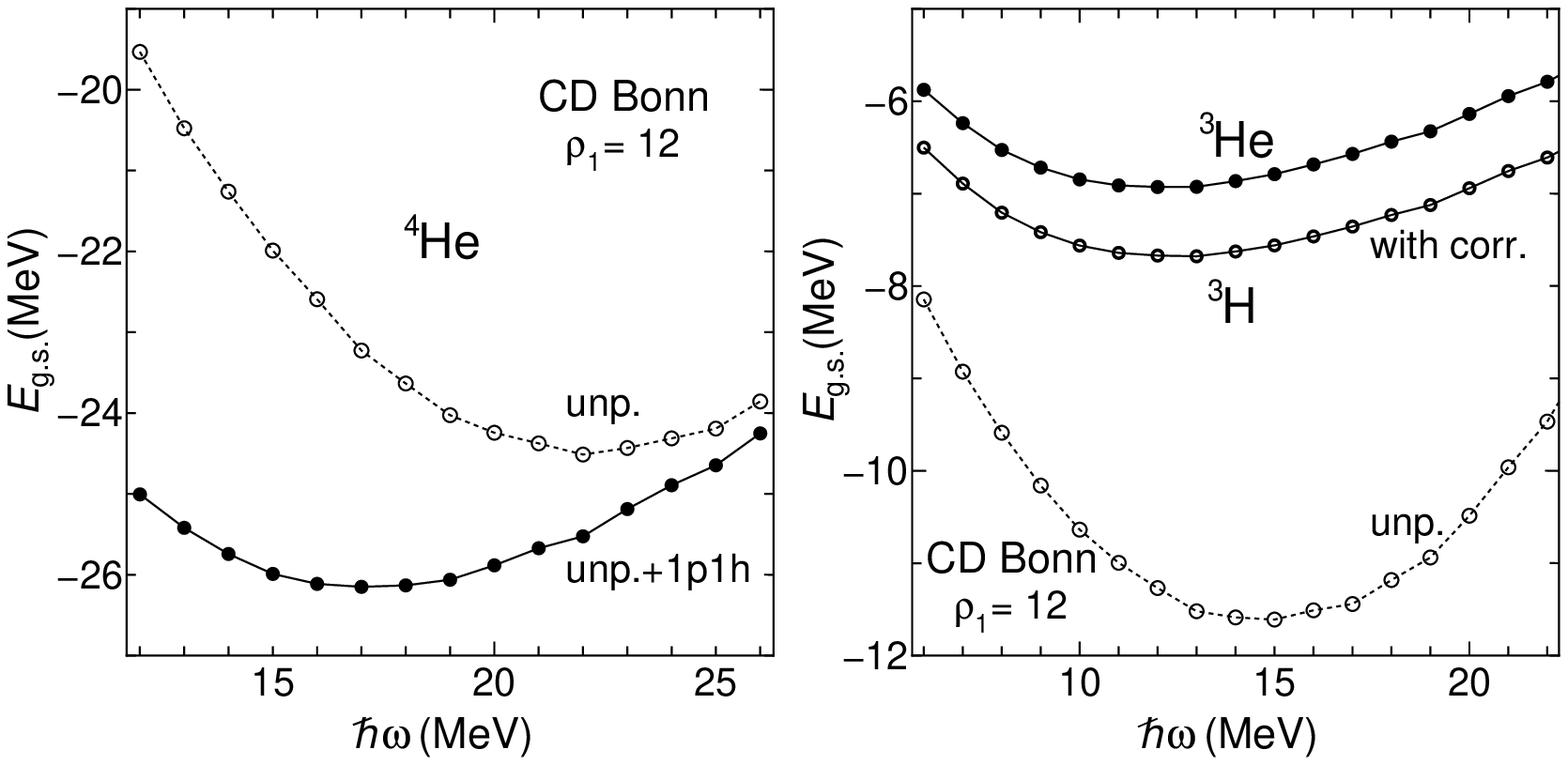}
\caption{\label{fig:He4_H3_He3_gs} The $\hbar \Omega$ dependence of calculated
ground-state energies of $^{4}$He, $^{3}$H, and $^{3}$He for $\rho _{1}=12$.
The CD-Bonn potential is employed.
}
\end{figure}

\begin{table}[t]
\caption{\label{tab:comparison} Comparison of the ground-state energies
of $^{4}$He, $^{3}$H, and $^{3}$He in the present approximation
for $\rho _{1}=12$ with those
in the NCSM calculations and the experimental values.
The CD-Bonn potential is commonly used in the calculations.
The experimental values are taken from Ref.~\cite{Audi93}.
All energies are in MeV.
}
\begin{ruledtabular}
    \begin{tabular}{cccc}
                      &    UMOA    &    NCSM    &   Expt.     \\ \hline
 $^{3}$He             & $  -6.93 $ & $  -7.25 $ & $  -7.72 $ \\
 $^{3}$H              & $  -7.68 $ & $  -8.00 $ & $  -8.48 $ \\
 $^{4}$He             & $ -26.15 $ & $ -26.30 $ & $ -28.30 $ \\
 $E_{\rm g.s.}$($^{3}$He)$-E_{\rm g.s.}$($^{3}$H)& $   0.75 $ & $   0.75 $ & $   0.76 $ \\
    \end{tabular}
\end{ruledtabular}
\end{table}

In Fig.~\ref{fig:He4_H3_He3_gs}, the $\hbar \Omega$ dependence of
calculated ground-state energies of $^{4}$He, $^{3}$H, and $^{3}$He
is shown for $\rho _{1}=12$ using the CD-Bonn potential.
The expression of the unperturbed ground-state energy is given
in Eq.~(\ref{eq:gs_energy}).
It is noted that the formulae for calculating the unperturbed
ground-state energies of $^{4}$He, $^{3}$H and $^{3}$He are the same.
In these cases, only the 0$s_{1/2}$ states of the proton and neutron are regarded
as the hole states.
However, the results of the unperturbed energies are different between
$^{4}$He and $^{3}$H ($^{4}$He and $^{3}$He) because of the $A$ dependence
of the Hamiltonian as given in Eq.~(\ref{eq:H_2nd}).
The expression of the ground-state energy of $^{4}$He with the $1p1h$
correction corresponds to Eq.~(\ref{eq:BE_O16}),
and that for $^{3}$H and $^{3}$He with the $1p2h$ effect is similar to
Eq.~(\ref{eq:BE_O15}).
We see that although the calculated results of
the unperturbed ground-state energies of $^{3}$H and $^{3}$He are the same,
the energies with the $1p2h$ effect are different because of the
charge difference.

In Table~\ref{tab:comparison}, the calculated ground-state energies of
$^{4}$He, $^{3}$H, and $^{3}$He with
the corrections for the optimal values of $\hbar \Omega$
which can be determined from Fig.~\ref{fig:He4_H3_He3_gs} are tabulated
together with the results of the no-core shell model
(NCSM)~\cite{Navratil00,Navratil01}
and the experimental values.
It has been shown that the NCSM results agree well with the results obtained
by accurate methods for few-nucleon systems
such as the Faddeev-Yakubovsky calculation~\cite{Kamada01}.
It is seen that our results are less bound by several hundred keV
than the NCSM results.
In the present approach, higher-order many-body correlations such as the
three-body cluster terms are not taken into account.
The evaluation of the higher-order many-body correlations would gain more
binding energy.
We may say, however, that our result of the charge dependence
in the relative energy between $^{3}$H and $^{3}$He is in good agreement with
the NCSM result and also the experimental value.

This kind of agreement of charge dependence can also be seen
in the results for $^{15}$N and $^{15}$O as shown in Tables~\ref{tab:N15_sp}
and \ref{tab:O15_sp}.
The experimental energy difference of the ground states between
$^{15}$N and $^{15}$O is $3.53$ MeV.
Our result of the energy difference of the single-particle energies
for the 0$p_{1/2}$ orbits between $^{15}$N and $^{15}$O
is $3.63$ MeV for the CD~Bonn.
One can see that the results for the other potentials also agree well with the
experimental value.
Thus, we may say that the effect of the Coulomb force for
the $pp$ channel is correctly treated in our particle-basis formalism.
The Coulomb force effect is also discussed in the next subsection for
$^{17}$F and $^{17}$O.

\subsection{\label{sec:F17_O17}$^{17}$F and $^{17}$O}

\begin{figure}[t]
\includegraphics[width=.365\textheight]{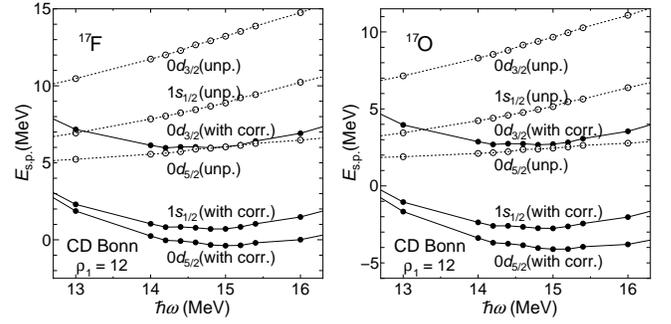}
\caption{\label{fig:F17_O17_hw} The $\hbar \Omega$ dependence of calculated
single-particle energies for $\rho _{1}=12$ in $^{17}$F and $^{17}$O.
The CD-Bonn potential is employed.
}
\end{figure}

Figure \ref{fig:F17_O17_hw} shows the $\hbar \Omega$ dependence
of calculated single-particle energies for $\rho _{1}=12$
for the 1$s$ and 0$d$ states
in $^{17}$F and $^{17}$O with the CD-Bonn potential.
The unperturbed energy and the energy with the $2p1h$ correction
are displayed separately.
The definition of the single-particle energy with the correction
is given in Eq.~(\ref{eq:SPE_P}).
We see that all the unperturbed energies are rather unbound
and considerably vary at around the typical $\hbar \Omega =14$ MeV.
However, some single-particle states become bound at the energy minimum points
by taking account of the corrections.
It should be noted that the magnitudes of the spin-orbit splitting
with the $2p1h$ effect for the 0$d$ states
are not very different from those for the unperturbed part at around
$\hbar \Omega = 14$ MeV.
This tendency differs from the case of the deeply bound hole states for which
the $1p2h$ effect plays an important role to enlarge the spin-orbit splittings
as shown in Fig.~\ref{fig:N15_O15_hw}.

\begin{figure}[t]
\includegraphics[width=.365\textheight]{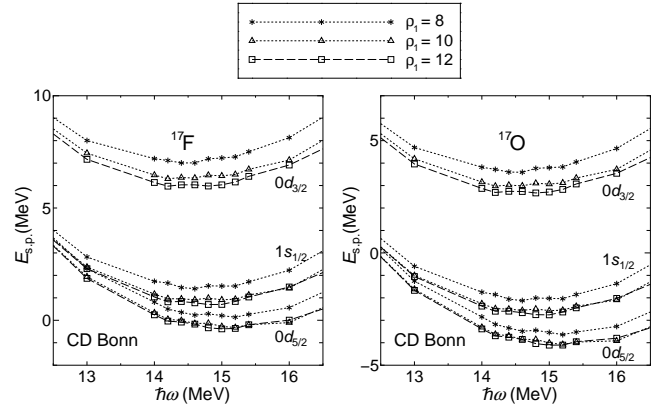}
\caption{\label{fig:F17_O17_hw_rho1} The $\hbar \Omega$ and $\rho _{1}$
dependences
of calculated single-particle energies with the $2p1h$ effect
in $^{17}$F and $^{17}$O
for the CD-Bonn potential.
}
\end{figure}

In Fig.~\ref{fig:F17_O17_hw_rho1}, we show the $\hbar \Omega$ and $\rho _{1}$
dependences of the single-particle energies with the
$2p1h$ effect.
We see that the $\rho _{1}$ dependence for the 0$d_{5/2}$ and 1$s_{1/2}$
states shows the good convergence at $\rho _{1}=12$.
On the other hand, the results for the 0$d_{3/2}$ states do not necessarily
converge at $\rho _{1}=12$.
Since the 0$d_{3/2}$ states of the proton and neutron are highly unbound,
it would be necessary to take a larger value of $\rho _{1}$
in order to obtain the convergent results.
In the present study, however, we employ the values for $\rho _{1}=12$
as the final results of the single-particle energies in $^{17}$F and $^{17}$O.

\begin{figure}[t]
\includegraphics[width=.330\textheight]{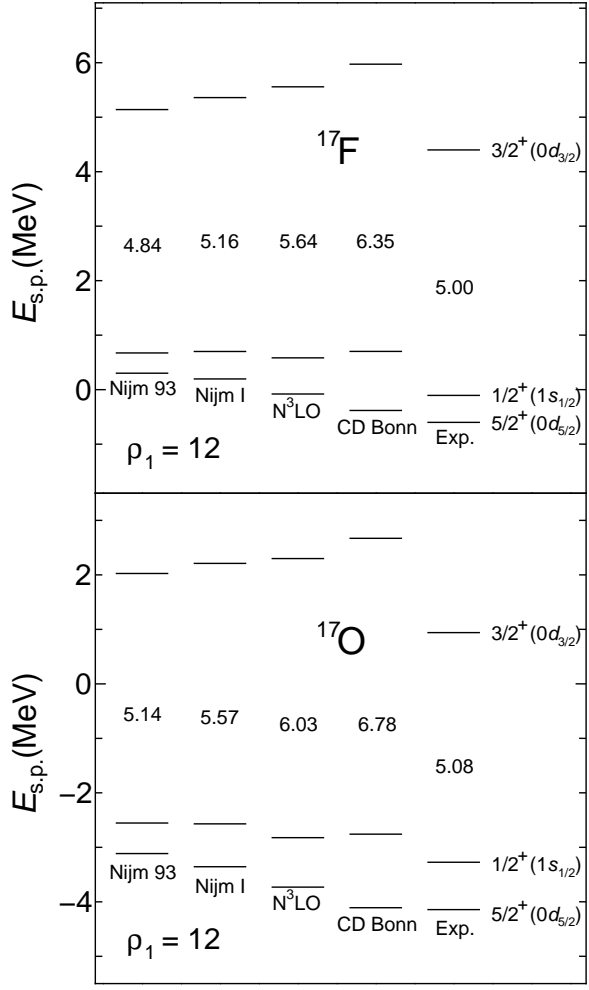}
\caption{\label{fig:F17_O17_sp}  The calculated single-particle energies
with the $2p1h$ effect for $\rho _{1}=12$ in $^{17}$F and $^{17}$O
for the CD-Bonn potential.
The values of the spin-orbit splitting are also shown.
In these calculated values, the optimal values of $\hbar \Omega$
for each single-particle state and interaction are employed.
}
\end{figure}

Shown in Fig.~\ref{fig:F17_O17_sp} are the final results of
the single-particle energies with the $2p1h$ effect
in $^{17}$F and $^{17}$O for the four potentials with the values of
the spin-orbit splitting energy.
The optimal values of $\hbar \Omega$ are employed for the results.
On the whole, the calculated spin-orbit energies are larger
than the experimental values in contrast to the hole state case.
We may say, however, that
the calculated results become somewhat smaller
if we take a larger value of $\rho _{1}$, because the 0$d_{3/2}$
states are lowered as suggested in Fig.~\ref{fig:F17_O17_hw_rho1}.
The calculated results for the ground 0$d_{5/2}$
states agree fairly well with the experimental values.
However, in our preliminary estimation,
the three-body cluster effect for the particle
state shows a repulsive contribution significantly
to the single-particle energy,
while that for the hole state is essentially small, as far as
only the two-body interaction is employed.

In Tables~\ref{tab:F17_sp} and \ref{tab:O17_sp}, the final results of the
single-particle energies shown in Fig.~\ref{fig:F17_O17_sp} are tabulated
together with the spin-orbit splitting energies for the 0$d$ states and
the energy differences between the 1$s_{1/2}$ and 0$d_{5/2}$ states.
The values for the typical $\hbar \Omega=14$ MeV are also displayed
in parentheses for reference.
We may say that our results for the magnitudes of these two splittings
are not very different from the experimental values.
However, we should take account of the real three-body force and evaluate
higher-order many-body correction terms to obtain more reliable results.
This kind of study is in progress.

We here discuss effects of the Coulomb force.
The experimental mass difference between $^{17}$F and $^{17}$O
is $3.54$ MeV.
The calculated results lie between $3.41$ and $3.73$ MeV for the
four potentials in the case of the optimal $\hbar \Omega$
and between $3.50$ and $3.63$ MeV in the case of $\hbar \Omega=14$ MeV
as seen from the values in Tables~\ref{tab:F17_sp} and \ref{tab:O17_sp}.
These calculated values are in good agreement with the experimental value.
This kind of agreement has been shown also for the hole states.
Furthermore, another effect of the Coulomb force appears in the particle
states.
The experimental 1$s_{1/2}$ states of the proton and neutron lie
above the 0$d_{5/2}$ states by $0.49$ and $0.87$ MeV in energy, respectively.
Thus, the 1$s_{1/2}$ state in $^{17}$F is close to the 0$d_{5/2}$ state
by $0.38$ MeV than in $^{17}$O.
This effect is known as the Thomas-Ehrman shift
due to the Coulomb force~\cite{Ehrman51,Thomas52,Aoyama98,Ogawa99}.
In our results, the magnitudes of the shift are from $0.19$ to $0.29$ MeV for
the cases of the optimal $\hbar \Omega$ and from $0.22$ to $0.23$ MeV for
$\hbar \Omega=14$ MeV, depending on the interactions employed.
In the latter case the results hardly depend on the potentials,
because the unperturbed 1$s_{1/2}$ h.o. wave functions are the same
for all the cases using the four interactions,
and thus the Coulomb force works equally in the calculations.
Although some discrepancies
between the experimental and calculated values are seen,
we may say that the Thomas-Ehrman effect can be observed in our results.

\begin{table}[t]
\caption{\label{tab:F17_sp} The calculated single-particle energies
with the $2p1h$ effect for $\rho _{1}=12$ in $^{17}$F.
The values of the spin-orbit splitting energy
$\Delta E_{ls}(0d)=E_{\rm s.p.}(0d_{3/2})-E_{\rm s.p.}(0d_{5/2})$
and the energy differences between the 1$s_{1/2}$ and 0$d_{5/2}$ states
$\Delta E_{sd}=E_{\rm s.p.}(1s_{1/2})-E_{\rm s.p.}(0d_{5/2})$
are also tabulated.
In these calculated values, the optimal values of $\hbar \Omega$
for each single-particle state and interaction are employed.
The results for $\hbar \Omega =14$ MeV are also shown in parentheses.
The experimental values are taken from Ref.~\cite{Tilley93}.
All energies are in MeV.
}
\begin{ruledtabular}
    \begin{tabular}{cccccc}
 $^{17}$F              & Nijm 93  &   Nijm I &   N$^{3}$LO  &  CD Bonn  &    Expt.   \\ \hline
 $ 3/2^{+}(0d_{3/2}) $ & $ 5.14 $ & $ 5.36 $ & $  5.56 $ & $  5.97 $ & $  4.40 $ \\
                       & $(5.50)$ & $(5.66)$ & $ (5.43)$ & $ (5.63)$ &  \\
 $ 1/2^{+}(1s_{1/2}) $ & $ 0.67 $ & $ 0.70 $ & $  0.58 $ & $  0.70 $ & $ -0.11 $ \\
                       & $(0.84)$ & $(0.83)$ & $ (0.52)$ & $ (0.54)$ &  \\
 $ 5/2^{+}(0d_{5/2}) $ & $ 0.30 $ & $ 0.20 $ & $ -0.08 $ & $ -0.38 $ & $ -0.60 $ \\
                       & $(0.22)$ & $(0.21)$ & $(-0.05)$ & $(-0.26)$ &  \\
 $ \Delta E_{sd}     $ & $ 0.37 $ & $ 0.50 $ & $  0.66 $ & $  1.08 $ & $  0.49 $ \\
                       & $(0.62)$ & $(0.62)$ & $ (0.57)$ & $ (0.80)$ &  \\
 $ \Delta E_{ls}(0d) $ & $ 4.84 $ & $ 5.16 $ & $  5.64 $ & $  6.35 $ & $  5.00 $ \\
                       & $(5.28)$ & $(5.45)$ & $ (5.48)$ & $ (5.89)$ &  \\
    \end{tabular}
\end{ruledtabular}
\end{table}

\begin{table}[t]
\caption{\label{tab:O17_sp}  Same as Table \ref{tab:F17_sp}, except for $^{17}$O.
}
\begin{ruledtabular}
    \begin{tabular}{cccccc}
 $^{17}$O              &   Nijm 93  &   Nijm I  &  N$^{3}$LO   &   CD Bonn  &    Expt.   \\ \hline
 $ 3/2^{+}(0d_{3/2}) $ & $  2.03 $ & $  2.21 $ & $  2.30 $ & $  2.67 $ & $  0.94 $ \\
                       & $ (2.33)$ & $ (2.47)$ & $ (2.19)$ & $ (2.37)$ &  \\
 $ 1/2^{+}(1s_{1/2}) $ & $ -2.55 $ & $ -2.57 $ & $ -2.82 $ & $ -2.76 $ & $ -3.27 $ \\
                       & $(-2.43)$ & $(-2.49)$ & $(-2.86)$ & $(-2.87)$ &  \\
 $ 5/2^{+}(0d_{5/2}) $ & $ -3.11 $ & $ -3.36 $ & $ -3.73 $ & $ -4.11 $ & $ -4.14 $ \\
                       & $(-3.28)$ & $(-3.33)$ & $(-3.66)$ & $(-3.89)$ &  \\
 $ \Delta E_{sd}     $ & $  0.56 $ & $  0.79 $ & $  0.91 $ & $  1.35 $ & $  0.87 $ \\
                       & $ (0.85)$ & $ (0.84)$ & $ (0.80)$ & $ (1.02)$ &  \\
 $ \Delta E_{ls}(0d) $ & $  5.14 $ & $  5.57 $ & $  6.03 $ & $  6.78 $ & $  5.08 $ \\
                       & $ (5.61)$ & $ (5.80)$ & $ (5.85)$ & $ (6.26)$ &  \\
    \end{tabular}
\end{ruledtabular}
\end{table}

\section{\label{sec:summary}Summary and conclusions}

The method for calculating the ground-state energy and single-particle energy
has been developed within the framework of
the unitary-model-operator approach (UMOA).
The expressions for the numerical calculation have been recast
from the isospin basis to the particle one for the charge-dependent
structure calculation.
We have applied the UMOA to $^{16}$O, $^{15}$N, $^{15}$O, $^{17}$F,
and $^{17}$O employing modern nucleon-nucleon interactions,
such as the Nijm-93, Nijm-I, the CD-Bonn, and the N$^{3}$LO potentials
which have charge dependence.
The Coulomb force has been also used for the $pp$ channel.
In order to obtain the final results,
we have searched for the optimal values of $\hbar \Omega$ and
the values of $\rho _{1}$ for which the calculated results
almost converge.

The accuracy of the approximation in the present method has been investigated
by calculating the ground-state energies of $^{4}$He, $^{3}$H, and $^{3}$He
and comparing the present results with
the accurate no-core shell-model (NCSM) results.
We have found that the energy differences between the NCSM and our results
for these systems are several hundred keV for the CD-Bonn potential.
As for the energy difference between $^{3}$H and $^{3}$He,
our result agrees well with the NCSM result and the experimental value.

The good agreement for charge dependence between the present results
and the experimental values is observed also in the differences
in the ground-state energies between $^{15}$N and $^{15}$O,
and $^{17}$F and $^{17}$O.
The effect of charge dependence is also seen in the Thomas-Ehrman shift for
the 1$s_{1/2}$ states in $^{17}$F and $^{17}$O.

We have shown that the calculated spin-orbit splittings for the 0$p$ hole
states are enlarged significantly by taking the $1p2h$ effect into account,
and become close to the experimental value.
On the other hand, the influence of the inclusion of the $2p1h$ effect
on the spin-orbit splittings for the 0$d$ particle states is rather small.
On the whole, the calculated spin-orbit splittings for the hole and particle
states in nuclei around $^{16}$O are not very different from the experimental
values though the results somewhat depend on the interactions employed.

In the present work, higher-order many-body correlations such as
the three-body cluster terms are not evaluated.
In addition, the real three-body force is not included in the calculations.
We should take account of these effects for a deeper
understanding of the nuclear structure.

By virtue of the extension of the calculation method to the particle basis,
the present method can be applied to proton- or neutron-rich nuclei
in the same manner.
The mechanism of the variation of magic numbers near the drip lines
may be clarified from a microscopic point of view.
The study of neutron-rich nuclei around $^{24}$O is in progress.
Results for these systems will be reported elsewhere in the near future.

\begin{acknowledgments}
The authors are grateful to D. R. Entem for providing us
the charge-dependent N$^{3}$LO potential code in reply to our request.
One of the authors (S.~F.)
acknowledges the Special Postdoctoral Researchers Program of RIKEN.
This work is supported by a Grant-in-Aid for Scientific Research (C) from
Japan Society for the Promotion of Science (JSPS) (No. 15540280).
\end{acknowledgments}

\end{document}